\documentclass[%
twocolumn,
superscriptaddress,
floatfix,
nofootinbib,
 amsmath,amssymb,
 aps, 
pre,
]{revtex4-2}

\usepackage{graphicx}% Include figure files
\usepackage{dcolumn}% Align table columns on decimal point
\usepackage{bm,xcolor,ulem}% bold math
\usepackage{amsmath,amssymb}
\usepackage{xcolor}

\providecommand{\prl}{Phys. Rev. Lett.}
\providecommand{\pre}{Phys. Rev. E}
\providecommand{\mnras}{Mon. Not. R. Astron. Soc.}
\providecommand{\nat}{Nature}

\usepackage{makecell}
\usepackage{booktabs} % For \toprule, \midrule, \bottomrule

\begin{document}

\preprint{APS/123-QED}

\title{Stress-enhanced diffusion in Yukawa crystals}

\author{M. E.\ Caplan}
\email{mecapl1@ilstu.edu}
\affiliation{Department of Physics, Illinois State University, Normal, Illinois 61761, USA}
\author{A. I.\ Chugunov}
%\email{andr.astro@mail.ioffe.ru}
\affiliation{Ioffe Institute, Politekhnicheskaya 26, 194021 Saint Petersburg, Russia}
\author{R. F.\ Serrano}
\affiliation{Department of Physics, Illinois State University, Normal, Illinois 61761, USA}

\date{\today}

\begin{abstract}
Self-diffusion in crystallized Coulomb (Yukawa) plasmas proceeds by the thermally activated formation of vacancy--interstitial pairs and the migration of these defects through the lattice, commonly known as Frenkel pairs in the materials literature. Molecular dynamics has recently resolved an Eyring law for diffusion in unstrained crystals, which we generalize to deformed crystals in this work. We simulate sheared Yukawa crystals and show that a static elastic strain exponentially enhances diffusion due to the additional elastic energy stored in the region that participates in Frenkel pair formation relative to the unstrained crystal:
$D(\varepsilon)=D_0(\Gamma^{*})\exp[V^{*}\sigma/k_{\rm B}T]$, with $\sigma=\tfrac12\mu\varepsilon^{2}$ the elastic energy density stored by the shear strain $\varepsilon$ and $V^{*}$ the activation volume. We find a single activation volume, $N_{\rm act} = n_iV^{*}\simeq31$ suffices for typical astrophysical screenings, with a reduction at higher screening. This suggests Frenkel pair formation requires only a brief coordination of the nearest neighbors at a thermal maximum in oscillation. Stress-enhanced diffusion supplies a microphysical mechanism for creep in strained neutron star crusts and crystallizing white dwarf cores.
\end{abstract}

\maketitle

\section{Introduction}

Strongly coupled Yukawa plasmas of fully ionized nuclei crystallize in the interiors of white dwarfs and in the crusts of neutron stars, as well as the dusty plasmas in laboratory
\cite{kirzhnits1960internal,Salpeter1961,vanHorn1968,Ruderman1968,Thomas1994,caplan2017colloquium,Saumon2022CurrentChallenges}. Self-diffusion in the solid phase therefore controls length and timescales for processes including element separation in white dwarfs and accreting neutron stars \cite{Mckinven2016,Baiko2024} and especially rheology (creep, plasticity, and failure) of magnetar crusts loaded by magnetic stresses \cite{Thompson2017,ChugunovHorowitz2010}. Recent molecular dynamics (MD) simulations showed that diffusion in Coulomb crystals proceeds by the spontaneous formation of vacancy--interstitial (Frenkel) pairs and the subsequent migration of the defects through the lattice \cite{Caplan2024}. That work also demonstrated a universal scaling relative to the melting temperature, ${\Gamma^{*}\equiv\Gamma/\Gamma_{\rm M}(\kappa)}$ with $\Gamma \propto1/T$, such that the dimensionless diffusion coefficient is independent of plasma screening parameter $\kappa$ and follows an Eyring law ${D^{*}=(\alpha/\Gamma^{*})e^{-\beta\Gamma^{*}}}$, made dimensionless by $D^{*}=D/(\omega_p a_i^{2})$ using the plasma frequency $\omega_p=(4\pi e^{2}Z^{2}n_i/m)^{1/2}$ and ion sphere radius $a_i=(4\pi n_i/3)^{-1/3}$. Here $Z$, $m$, $n_i$ are the charge, mass, and number density of the nuclei, and $e$ is the elementary charge.

Astrophysical neutron star crusts, however, are elastically stressed by magnetic fields in magnetars, by accretion, and by rotation or spin down, up to breaking strains of order
$0.1$ \cite{horowitz2009breaking,ChugunovHorowitz2010,Hoffman,BK17,Baiko2018}. Because Frenkel pair formation is thermally activated, elastic energy stored in the lattice should reduce the effective
barrier and enhance diffusion, by analogy to stress-assisted creep in ordinary terrestrial solids. In this work we study this effect with MD simulations of statically sheared body-centered cubic (bcc) Yukawa crystals and show that the stress-enhancement is a simple Arrhenius law with only one parameter -- the activation volume.

In Sec. \ref{sec:theory} we introduce our theory of stress-enhanced diffusion and derive the dimensionless argument of an exponential Arrhenius enhancement. In Sec. \ref{sec:md} we describe our molecular dynamics formalism, and in Sec. \ref{sec:results} we present results and fits.

\section{Theory of Stress-Enhanced Diffusion}\label{sec:theory}

Diffusion by thermally activated Frenkel pair formation should obey an Arrhenius law, 
$D\propto e^{-\Delta U/k_{\rm B}T}$, where $\Delta U$ is the free-energy barrier to nucleate the defect \cite{Caplan2024,Hughto2011Diffusion}. In the unstrained crystal this rate is the universal Eyring law of Caplan \& Yaacoub ~\citep[hereafter CY24]{Caplan2024},
\begin{equation}
  D_0(\Gamma^{*}) = \frac{\alpha}{\Gamma^{*}}\,e^{-\beta\Gamma^{*}},
  \label{eq:cy}
\end{equation}
\noindent and is a function of the thermodynamic state alone (through $\Gamma^{*}$, hence $T$ and $\kappa$). Our hypothesis is that a static elastic strain lowers the barrier by the elastic energy stored in the region that participates in Frenkel pair formation, \textit{i.e.} an activation volume $V^{*}$ holding elastic energy $V^{*}\sigma$, with $\sigma$ the stored elastic energy density. We therefore generalize Eq.~(\ref{eq:cy}) to strained crystals by multiplying by a single factor, the Boltzmann factor of that stored energy,
\begin{equation}
  \,D(\Gamma^{*},\varepsilon) \;=\; D_0(\Gamma^{*})\,
       \exp\!\left[\frac{V^{*}\sigma}{k_{\rm B}T}\right]\,,
  \qquad \sigma=\tfrac12\,\mu\,\varepsilon^{2}.
  \label{eq:master}
\end{equation}
\noindent with strain $\varepsilon$ and elastic modulus $\mu$. Observe that this does not require any other activation energy or free parameters beyond the activation volume. 

The temperature follows from the definition
of the coupling parameter,
\begin{equation}
  k_{\rm B}T = \frac{Z^{2}e^{2}}{a_i\,\Gamma},
  \qquad \Gamma=\Gamma^{*}\,\Gamma_{\rm M}(\kappa),
  \label{eq:kT}
\end{equation}
with the Yukawa crystal melt line
\begin{equation}
  \Gamma_{\rm M}(\kappa)=\Gamma_{\rm crit}\,\frac{e^{\bar\kappa}}{1+\bar\kappa+\bar\kappa^{2}/2},
  \qquad \bar\kappa=\left({4\pi}/{3}\right)^{1/3}\kappa ,
  \label{eq:melt}
\end{equation}
and $\Gamma_{\rm crit}=175.7$ \cite{vaulina2000scaling,Vaulina2002Universal,baiko2022ab}. A word on conventions is necessary here because two definitions of the screening parameter are in use. Throughout this work $\kappa\equiv a_i/\lambda$  ($\lambda$ is Yukawa screening length, i.e.
the interaction potential is $\propto e^{-r/\lambda}$).
Refs.~\cite{vaulina2000scaling,Vaulina2002Universal,Caplan2024} instead use $\kappa_{\rm CY}\equiv n_i^{-1/3}/\lambda=(4\pi/3)^{1/3}\kappa=\bar\kappa$, which is why $\bar\kappa$ rather than $\kappa$ appears in Eq.~(\ref{eq:melt}) and in the dust lattice wave frequency $\omega_{\rm DLW}=\omega_p\,e^{-\bar\kappa/2}(1+\bar\kappa+\bar\kappa^2/2)^{1/2}$ that sets the MD timestep. The screening lengths $\kappa_{\rm CY}=0.537$ and $1.074$ of Ref.~\cite{Caplan2024} are $\kappa=0.333$ and $0.666$ here. The elastic moduli below are tabulated in terms of $\kappa=a_i/\lambda$.

As we are primarily interested in shear deformation, the modulus is that conjugate to the imposed $xy$ shear, $\mu=C_{44}$, in dimensionless form $\tilde\mu=\tilde c_{44}$ in units of $n_i Z^{2}e^{2}/a_i$. We take the $T=0$ bcc value from the Yukawa fit of Kozhberov \cite{Kozhberov2022} [Eq.~(22), bcc Table~II],
\begin{equation}
  \tilde c_{44}(\kappa)=\frac{k_0+k_1\kappa^{2}+k_2\kappa^{4}}{1+k_3\kappa^{2}+k_4\kappa^{4}},
  \label{eq:c44}
\end{equation}
with $k_0=0.18277$, $k_1=-0.0113456$, $k_2=2.26914\times10^{-4}$, $k_3=0.162907$, $k_4=8.92591\times10^{-3}$. 
It gives $\tilde c_{44}=0.166$, $0.147$, and $0.0101$ at $\kappa=0.666$, $1$, and $4$. To this we add the classical thermal softening of the shear modulus, Eq.~(46) from Baiko \cite{Baiko2011},
\begin{equation}
  \tilde\mu(\kappa,\Gamma)=\tilde c_{44}(\kappa)-\frac{439^{1/3}}{\Gamma}\simeq\tilde c_{44}(\kappa)-\frac{7.60}{\Gamma},
  \label{eq:soften}
\end{equation}
which amounts to $\tilde\mu=0.128$, $0.114$, and $0.0082$ at $\Gamma^*=1.05$ for the three screenings used in Sec.~\ref{sec:results}. 

It is natural to measure $V^{*}$ in units of the ion-sphere volume
$4\pi a_i^{3}/3=1/n_i$ such that the activation number $N_{\rm act}\equiv n_i V^{*}$ counts the ions in the
activation region, and the argument of the added exponent takes the compact dimensionless form
\begin{equation}
  \frac{V^{*}\sigma}{k_{\rm B}T}
    = N_{\rm act}\,\frac{\sigma V_{\rm ion}}{k_{\rm B}T}
    = N_{\rm act}\cdot\tfrac12\,\tilde\mu\,\Gamma\,\varepsilon^{2},
  \label{eq:exponent}
\end{equation}

\begin{figure}
  \includegraphics[width=\columnwidth]{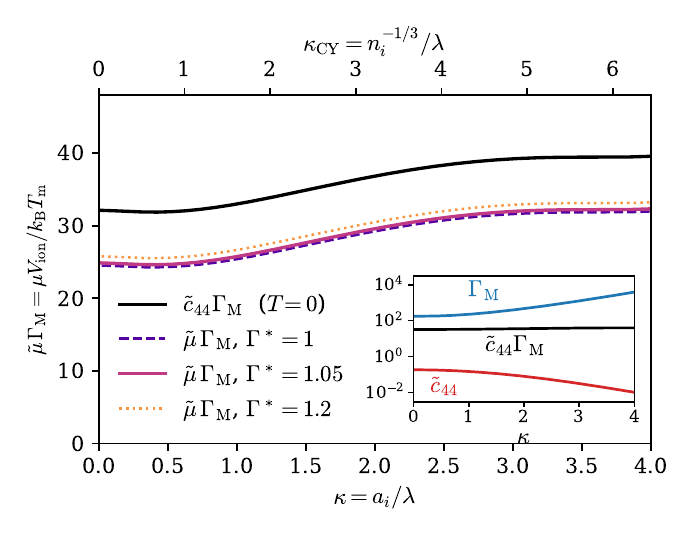}
  \caption{The product of $\tilde\mu\Gamma_{\rm M}$ is approximately constant across $\kappa$, but does vary by about a third between $\kappa<1$ and $\kappa>1$.}
  \label{fig:muGamma}
\end{figure}

\noindent Equation~(\ref{eq:exponent}) already suggests that the strain enhancement should be nearly independent of screening when expressed in terms of $\Gamma^*$. Writing $\tilde\mu\Gamma=(\tilde\mu\,\Gamma_{\rm M})\,\Gamma^*$, the product $\tilde\mu\,\Gamma_{\rm M}$ is approximately constant in $\kappa$. Observe that the melt line of Eq.~(\ref{eq:melt}) follows from the Lindemann criterion applied to the dust lattice wave frequency \cite{vaulina2000scaling,Vaulina2002Universal}, so $\Gamma_{\rm M}\propto 1/\tilde\mu$ to the extent that the Lindemann ratio is universal. We verify this with a plot of $\tilde\mu\Gamma_{\rm M}$ in Fig.~\ref{fig:muGamma}.

With Eqs.~(\ref{eq:melt}) and (\ref{eq:c44}), $\tilde c_{44}\Gamma_{\rm M}=31.9,\,32.1,\,33.0,$ and $39.6$ for $\kappa=0.333,\,0.666,\,1,$ and $4$, a $25\%$ variation over a factor of $20$ in $\Gamma_{\rm M}$ (with the thermal softening of Eq.~(\ref{eq:soften}) included, $\tilde\mu\Gamma$ at $\Gamma^*=1.05$ is $25.9$, $26.1$, $27.0$, and $33.9$). The elastic energy stored per ion in units of $k_{\rm B}T$ at fixed $\varepsilon$ and $\Gamma^*$ is therefore nearly the same for any screening, and the universality found for $D_0(\Gamma^*)$ in Ref.~\cite{Caplan2024} should extend to the sheared crystal provided $V^*$ itself has weak sensitivity to $\kappa$. 

As an ansatz, we propose that an activation number of about $3^3 \approx 30$ is likely, as Frenkel pair formation requires the coordination of a particle on its lattice site with its nearest neighbors at thermal maxima so that a vacancy--interstitial pair can form and separate. 
We now check this explicitly with molecular dynamics.

\section{Molecular dynamics}\label{sec:md}

Our molecular dynamics formalism is the same as in CY24 \cite{Caplan2024}, but now with a shear deformation.

Ions interact through the screened Coulomb (Yukawa) potential $V(r)=(Z^{2}e^{2}/r)\,e^{-r/\lambda}$. We simulate ions ($Z=10$, $A=20$) on an $M^{3}$ bcc lattice ($N=2M^3$) in a periodic box with LAMMPS \cite{LAMMPS}, with timestep $\omega_{\rm DLW}^{-1}/17$ and a Nose-Hoover thermostat. Two sets of simulations are used:
(1) a screening scan of $20^3$ cells ($N=16000$) at $\kappa\equiv a_i/\lambda=0.666$, $1$, and $4$ is run at fixed $\Gamma^*=1.05$;  and (2) a temperature scan of $16^3$ cells ($N=8192$) at $\kappa=0.666$ is run at $\Gamma^*=1.05$--$1.20$ with several random seeds per state point.
The pair potential is truncated at half the box length, $r_{\rm cut}=L/2$.

A shear strain $\varepsilon$ is imposed by tilting the cell in the $xy$ plane to transform the cubic simulation box into a parallelepiped, the deformation conjugate to the $c_{44}$ shear modulus. The deformation is held fixed while the crystal equilibrates. Two slabs of ions, single slabs of lattice planes ($\approx6\%$ of the volume for $M=16$, $5\%$ for $M=20$), at the $\pm y$ boundaries are frozen to anchor the strain. The frozen layers are also used in the zero strain runs for consistency. This is necessary because the formation of vacancy-interstitial pairs at astrophysically relevant screening lengths is only possible within about 10\% of the melting temperature in molecular dynamics simulations, but at such high temperature the breaking strain is significantly reduced
\cite{ChugunovHorowitz2010,Hoffman,Baiko2011}.
After equilibration, we calculate the mean squared displacement (MSD) of all ions and obtain the dimensionless diffusion coefficient $D^{*}=\mathrm{MSD}/(6t\,\omega_p a_i^{2})$ from a two-point slope between an early time, after the MSD has reached its vibrational plateau of $\sim0.1a_i^2$, and the end of the run. Here $\omega_p$ is the unscreened plasma frequency, as in Ref.~\cite{Caplan2024}; the MD timestep is $(17\,\omega_{\rm DLW})^{-1}$ with the screened dust lattice frequency $\omega_{\rm DLW}(\bar\kappa)$ of Sec.~\ref{sec:theory}, which at $\kappa=4$ differs from $\omega_p$ by a factor of $5$, so care is required when converting step counts to time. 

The frozen slabs impose the strain but their ions cannot participate in Frenkel pair cascades and they also suppress nucleation in the adjacent layer, while the MSD is averaged over all $N$ ions. With a box of $20^3$ unit cells, including these immobile ions may weight the MSD by a fraction $f_{\rm act}\approx1-4/20\approx 0.8$. For a smaller box of $16^3$, $f_{\rm act} \approx 0.75$. This correction is strain independent so it does not affect the signal in the strain response we are most interested in. Rather, it is effectively a prefactor rescaling the unstrained baseline $D_0(\Gamma^{*})$, which was reported in CY24 to have about 50\% uncertainty anyway. We are therefore unconcerned with this small correction.

\section{Results}\label{sec:results}

\begin{figure}[t!]
  \includegraphics[width=\columnwidth]{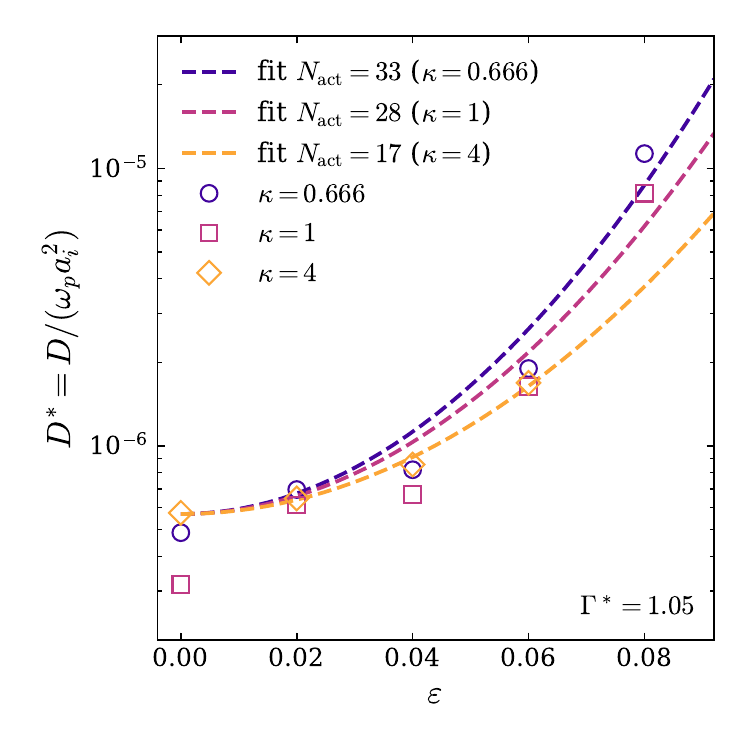}
  \caption{Diffusion coefficients of the sheared bcc Yukawa crystal grow like $\varepsilon^2$ at fixed $\Gamma^*=1.05$ for three screening lengths. We fit using Eq.~\ref{eq:master} ($\alpha=787$, $\beta=20$) with the exact form of the elastic energy exponent given by Eq.~\ref{eq:exponent}.} 
  \label{fig:kappa}
\end{figure}

We begin by testing the screening dependence in Fig.~\ref{fig:kappa}, and the $\varepsilon^2$ dependence is immediately apparent. These simulations were run for at least $10^7$ timesteps, with those at $\kappa=0.666$ run $4\times10^7$ timesteps. Figure~\ref{fig:kappa} shows $D^*(\varepsilon)$  for $\kappa=0.666$, $1$, and $4$ at fixed $\Gamma^*=1.05$; these simulations span a factor of 20 in $\Gamma$ (from $204$ to $4100$) due to the dependence of the melting temperature on $\kappa$. Several features stand out.

First, the unstrained diffusion coefficients ($\varepsilon=0$) at all three screenings agree with one another and with the universal law of Ref.~\cite{Caplan2024} at the few-$10^{-1}$ level. The agreement with CY24 is generally better at higher screening, as expected. From CY24, we know that Frenkel pair formation happens at a greater rate at higher screenings, while for lower screenings ($\kappa \lesssim 1$) simulation runs of this size only form a small number of defects.

Second, the strain enhancement is largely the same at $\kappa=0.666$ and $\kappa=1$, with only a modest reduction at larger screenings of $\kappa=4$. For $\kappa\lesssim1$, as expected from our arguments above, a constant $\tilde\mu \Gamma_M$ tends to suggest a constant activation number implied by the hypothesis of universal diffusion rates and an activation number of order $3^3$. This is perhaps the most useful key result of this work, namely, that $N_{\rm act} \approx 31\pm1$ holds across astrophysically relevant screenings. Notably, our highest screening (which also has the best defect formation statistics) has a slightly lower $N_{\rm act}=17$.

Note that due to the better activation statistics at higher strains the slope is set largely by the two highest strains. The simulation that would provide the most useful data point for checking the universality, $\varepsilon=0.08$ at $\kappa=4$, unfortunately yielded almost immediately, and so no useful information could be collected from it.

With the quadratic dependence implied by this figure, we now attempt to determine $N_{\rm act}$ precisely, which requires an ensemble of many seeds, and also allows us to confirm temperature independence. For this we use a smaller $16^3$ grid at $\kappa=0.666$. The smaller cell carries a known systematic
\cite{Hughto2011Diffusion}, which suppresses the unstrained baseline $D_0$ by a factor of order $3$ relative to Eq.~(\ref{eq:cy}).
However, the small size of the activation volume is exactly what allows us to tolerate this. Because $V^*$ corresponds to a sphere of radius $\approx3a_i$, far inside the cell, and because the shear is volume conserving, we expect the finite-size error to be almost the same at every $\varepsilon$ and the bias cancels in the ratio $D(\varepsilon)/D_0(\Gamma^*)$ from which $N_{\rm act}$ is measured. A good agreement of $N_{\rm act}$ for our $N=16000$ and $N=8192$ simulations supports this assumption. We may therefore accept an error in the prefactor $D_0$ in exchange for the statistics needed to obtain the exponent with less investment of computing. As we demonstrate below, our results strongly indicate that the reason for the finite size bias in the prefactor $D_0$ is the length of the diffusion chain, associated with the lifetime of the Frenkel pair, but not the probability of formation of such pair.

\begin{figure}[t!]
  \includegraphics[width=\columnwidth]{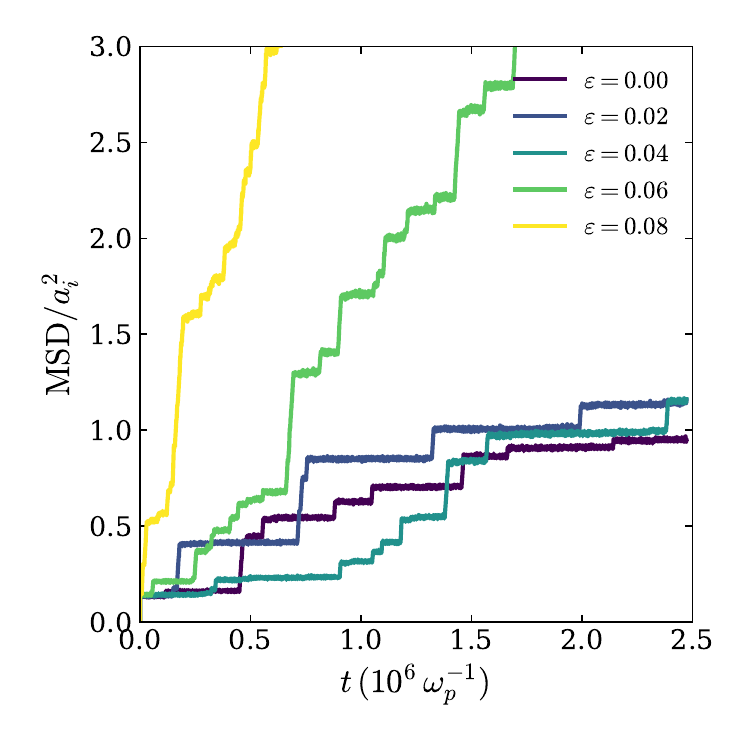}
  \caption{Fiducial mean squared displacements obtained at $\Gamma^* = 1.10$ show that the diffusion coefficient (mean slope) should have a strong dependence on strain, but only through the defect formation rate as the defect migration rate (instantaneous slope during activity) is strain independent.}
  \label{fig:fiducial}
\end{figure}

\begin{figure*}[t!]
  \includegraphics[width=\textwidth]{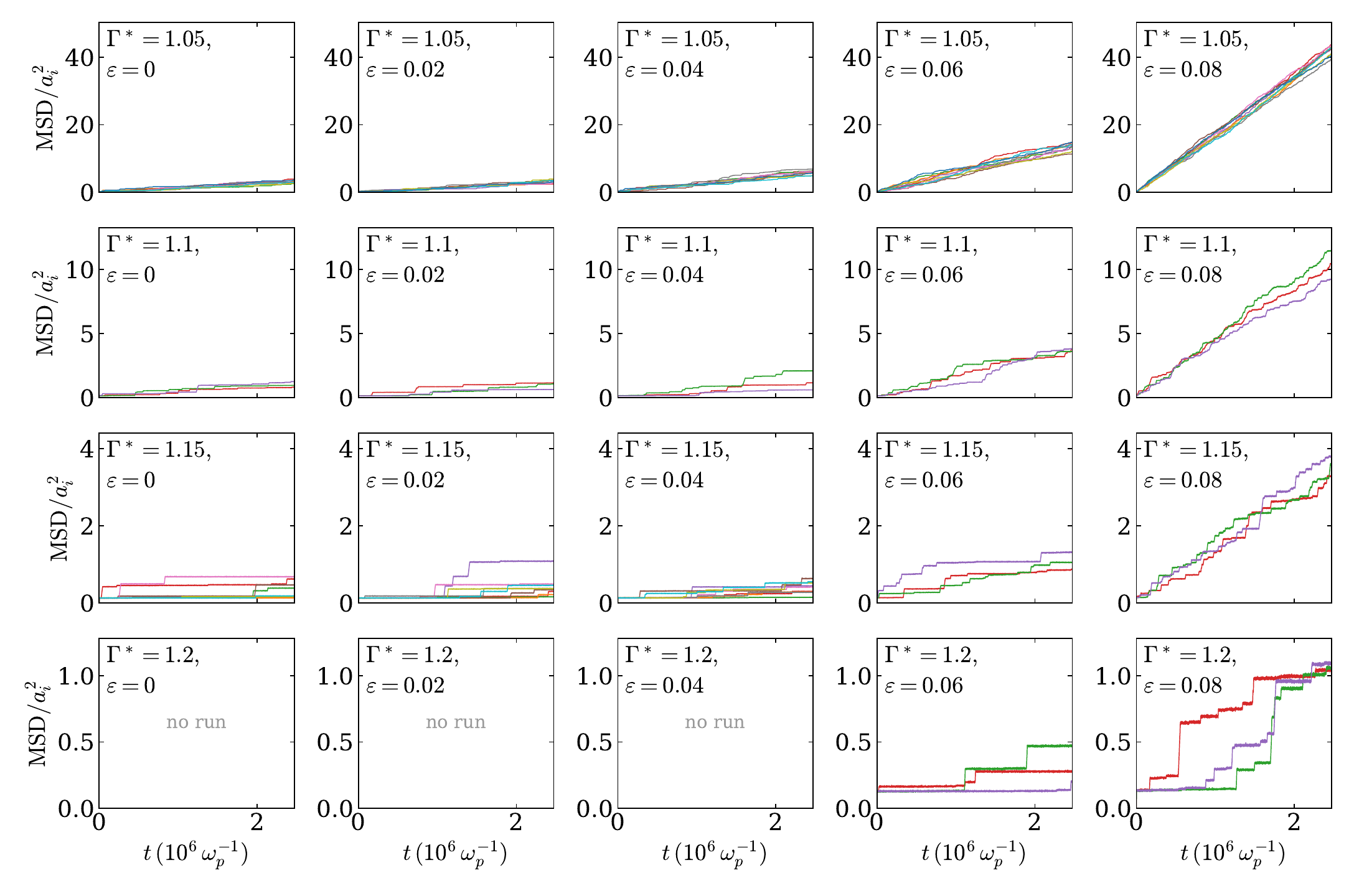}
  \caption{Time dependent mean squared displacements as a function of time for a grid of simulations at $\kappa=0.666$. The strong dependence on $\Gamma^*$ from CY24 is observed reading columns vertically, while a newly resolved and apparently quadratic dependence on shear strain $\varepsilon$ at fixed $\Gamma^*$ can be seen reading rows.}
  \label{fig:MSDt}
\end{figure*}

We present the MSD obtained from a fiducial set of simulations across five strains at $\Gamma^*=1.10$ in Fig. \ref{fig:fiducial}. The stochastic stair-step behavior is due to the formation of Frenkel pairs and their subsequent migration through the lattice which only returns to quiescence when the vacancy and interstitial encounter each other and annihilate. Two features are immediately apparent.
By inspection, the instantaneous slope of the MSD during these periods of activity show no dependence on $\varepsilon$; this is most easily seen in the bottom left at early times before the curves fan out. This suggests that the jump frequencies of the vacancies and interstitials are insensitive to the elastic energy of the lattice\footnote{Note that we are unable to separate species-wise contributions from the vacancy and interstitial to $D^*$, but there is no reason to suspect an enhancement in the migration of one would be exactly canceled by a suppression in the migration rate of the other.}. However, the duty cycle, the fraction of time when a defect is active, shows a clear dependence on shear strain. Equivalently, the number of `steps' in any given MSD curve appears to correlate with $\varepsilon^2$, supporting our hypothesis of a formation activation enhancement. At early times, the MSD jumps to about $0.1a_i^2$ within the first few timesteps due to thermal noise on the lattice site, which must be subtracted off when calculating $D^*$ later; this has been effectively handled in past work using a Heaviside cut-off in displacement to reject the thermal contribution of particles that have not moved at least one lattice site \cite{Hughto2011Diffusion,Caplan2024}. Also observe that the duty cycle at $\varepsilon = 0.08$ is so high that the first defect forms almost immediately after the simulation starts.

Given that the instantaneous slope of the MSD during a cascade is by inspection strain independent, we must conclude that the diffusion coefficient, which is proportional to a defect density multiplied by a defect hop rate, must be overwhelmingly dominated by the strain dependence of defect formation rate. In truth, we are able to fit a $\sim 10\%$ enhancement in the migration rate between $\varepsilon=0$ and 0.08, but this is again well within the kind of stochastic error we have due to the defect formation rate, and so we do not try to isolate it in more detail.

We extend our data set across a grid of $(\Gamma^*, \varepsilon)$, running multiple random seeds for each to average over the stochasticity. Our grid includes 107 simulations and are shown in Fig.~\ref{fig:MSDt}. Of these, 50 are at our hottest $\Gamma^*=1.05$ as this will provide the best statistics. Each panel corresponds to a unique $(\Gamma^*, \varepsilon)$ and includes all seeds, with rows corresponding to $\Gamma^*$ and columns correspond to each value of $\varepsilon$. At fixed $\Gamma^*$ we plot with identical vertical axes for easy comparison. 

\begin{figure*}[t]
  \includegraphics[width=\textwidth]{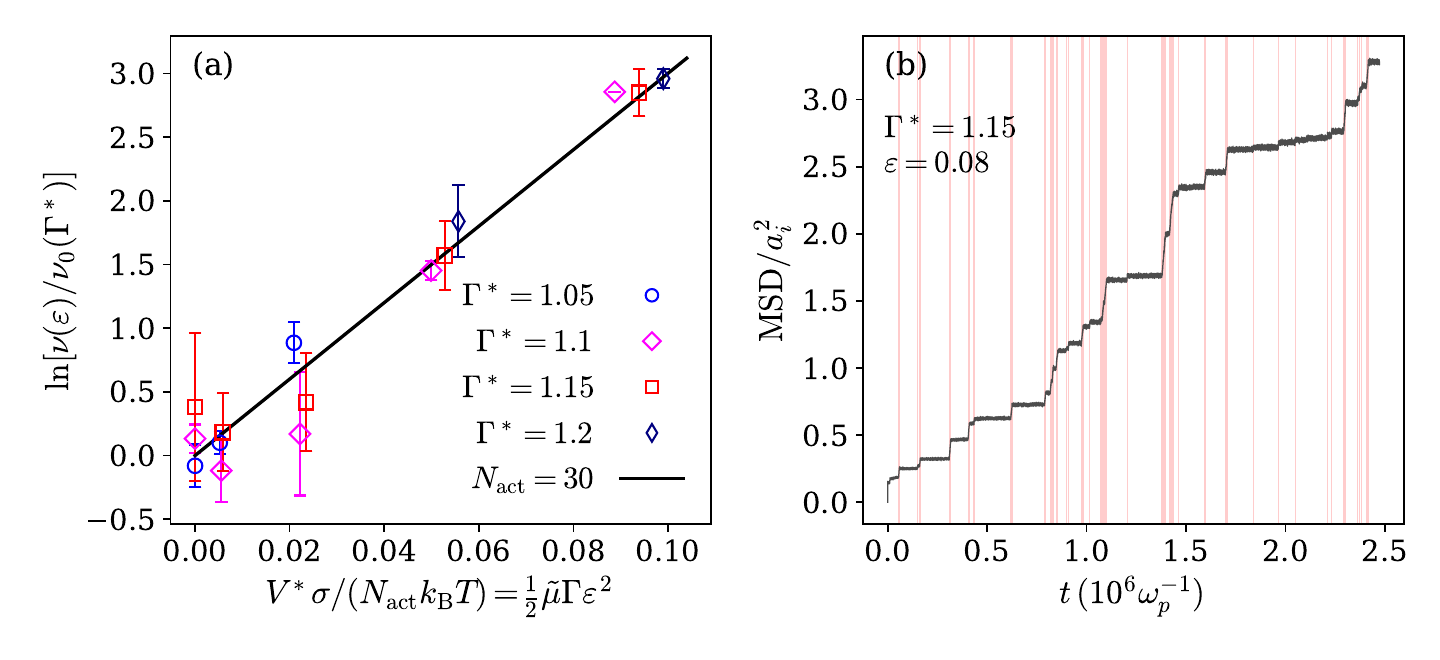}
  \caption{(a)~Cascade nucleation rates show a common activation number $N_{\rm act}$ across all temperatures and strains simulated. For each simulation the rate
  $\nu$ of resolved diffusive cascades is measured from the segmented
  ${\rm MSD}(t)$; by plotting $\ln[\nu(\varepsilon)/\nu_0(\Gamma^*)]$ against the elastic energy stored per ion,
  $\tfrac12\tilde\mu\,\Gamma\varepsilon^{2}$, collapses all couplings onto one
  line whose common slope is the activation number $N_{\rm act}$. This fit is independent from the $D^{*}(\varepsilon)$ fit below. (b) An example ${\rm MSD}(t)$ 
   at ($\Gamma^{*}=1.15$, $\varepsilon=0.08$) with the detected cascades shaded. }
  \label{fig:rate}
\end{figure*}

If a static strain affects only the formation of Frenkel pairs, then the rate $\nu$ at which cascades begin should carry the entire strain dependence with the same exponent as the diffusion coefficient -- an expression for $\nu$ follows immediately by analogy from Eq. (\ref{eq:master}).  It suffices to count the number of `steps' in the MSD and compare across strains, though these are admittedly made noisy by some very short lived pairs that recombine and annihilate. We count the cascades by first smoothing the MSD to reduce the thermal noise. From there a threshold on the instantaneous derivatives make it easy to identify intervals when defects are active, which is a reliable count of independent activation events when the duty cycle is low (such that a defect does not form when another is active). 
Thus, our analysis includes all cascades across all simulations where the duty cycle is less than 0.5;
this only required us to omit the simulations at $(\Gamma^*,\varepsilon)=(1.05, \,0.06)$ and $(1.05, \, 0.08)$.
One example of the counting for a representative MSD is shown in Fig.~\ref{fig:rate}(b); the algorithm counts about 31 events. It may incorrectly combine two that start and end in a short time period, and may miss some others that are very short lived. This analysis should therefore be considered heuristic and illustrative. 

We argue by analogy with Eq.~\eqref{eq:master} that $\nu$ obeys
\begin{equation}
  \,\nu(\Gamma^{*},\varepsilon) \;=\; \nu_0(\Gamma^{*})\,
       \exp\!\left[\frac{V^{*}\sigma}{k_{\rm B}T}\right].
  \label{eq:numaster}
\end{equation}
\noindent For each coupling $\Gamma^{*}$ we fit
\begin{equation}
  \ln\nu(\Gamma^{*},\varepsilon) \;=\; \ln\nu_{0}(\Gamma^{*}) \;+\;
  \tfrac{1}{2}N_{\rm act}\,\tilde\mu\,\Gamma\,\varepsilon^{2},
  \label{eq:rate_fit}
\end{equation}
where explicitly $\nu(\Gamma^{*},\varepsilon)$ is the cascade nucleation rate at coupling
$\Gamma^{*}$ and strain $\varepsilon$ and $\nu_{0}(\Gamma^{*})$ is the corresponding
unstrained nucleation rate, with  $\tfrac{1}{2}\tilde\mu\,\Gamma\,\varepsilon^{2}$ the dimensionless elastic energy stored per ion as before.
Equation~(\ref{eq:rate_fit}) has a free intercept
$\ln\nu_{0}(\Gamma^{*})$ per $\Gamma^*$. We use the fitted unstrained nucleation rate
at that $\Gamma^{*}$, determined from $\varepsilon$ at that $\Gamma^*$, rather than normalizing to the $\varepsilon=0$ measurement which has greater MD stochasticity. We fit a shared $N_{\rm act}$ to all four couplings.
Subtracting each coupling's own fitted intercept from both sides of
Eq.~(\ref{eq:rate_fit}) collapses the data onto one line through the origin,
\begin{equation}
  \ln\!\left[\frac{\nu(\varepsilon)}{\nu_{0}(\Gamma^{*})}\right]
  \;=\;
  \tfrac{1}{2}N_{\rm act}\,\tilde\mu\,\Gamma\,\varepsilon^{2}.
  \label{eq:rate_collapse}
\end{equation}

\noindent This is plotted in Fig.~\ref{fig:rate}(a). Observe that $\ln\nu$ rises linearly with the stored elastic energy per ion $\tfrac12\tilde\mu\Gamma\varepsilon^{2}$ and with the unstrained rate $\nu_0(\Gamma^{*})$ divided out all four couplings fall on a single line. From its slope we obtain an activation number $N_{\rm act} \approx 30$. We emphasize that this activation number is obtained entirely from counting the cascades that occur using no information about their magnitude (\textit{i.e.} the cascade duration or defect lifetime and migration frequency) which together set the diffusion coefficient. 

\begin{figure*}[t]
  \includegraphics[width=\textwidth]{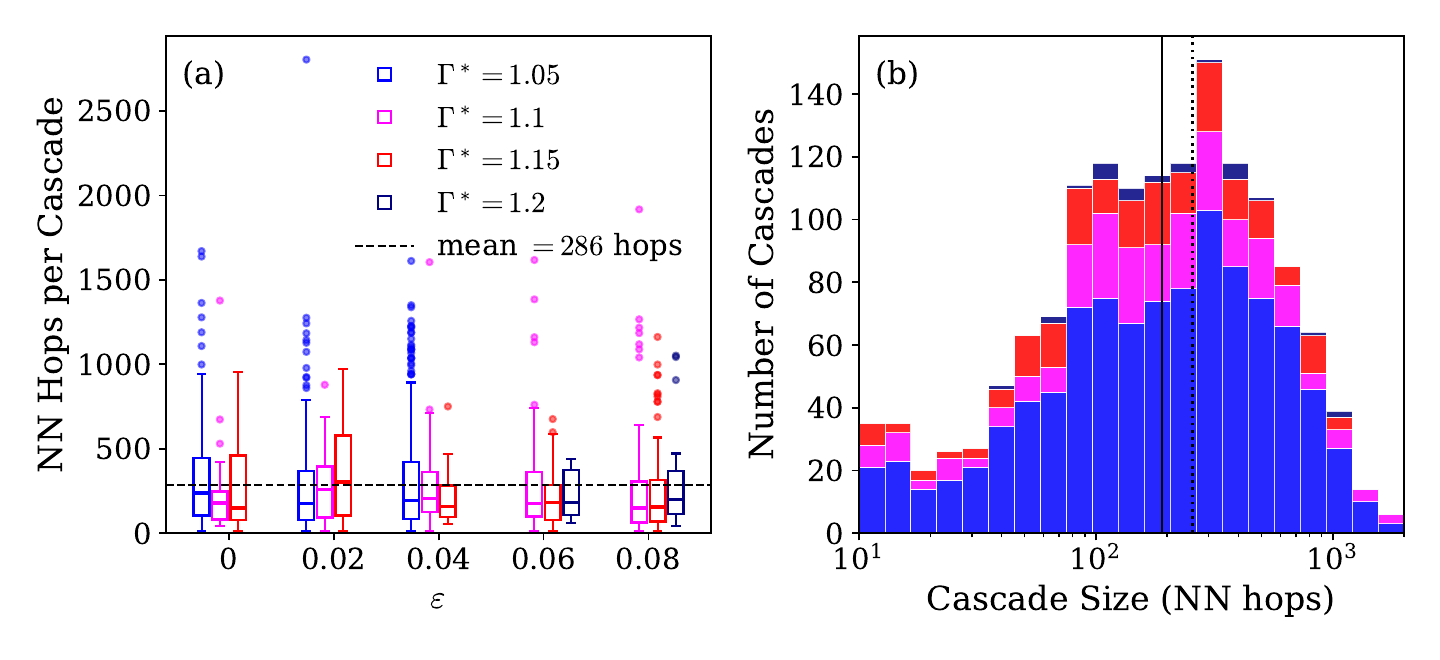}
  \caption{(a)~Mean cascade size (nearest-neighbor-hop equivalents,
  $N\,\Delta{\rm MSD}/d_{\rm nn}^{2}$) versus strain shows no obvious dependence on shear strain $\varepsilon$ or coupling $\Gamma^*$. (b)~Histogram
  of all individual cascade sizes (log scale), stacked and colored by
  $\Gamma^{*}$; each coupling contributes across the entire range with more or less the same
  shape. The distribution is broad and peaked near the diffusive box-crossing scale $M^{2}$ (dotted; $M=16$) with a tail of rare long
  excursions, as expected for a random walk to re-encounter in a finite cell.} \label{fig:size}
  
%\end{figure*}
%\begin{figure*}[]
  \includegraphics[width=\textwidth]{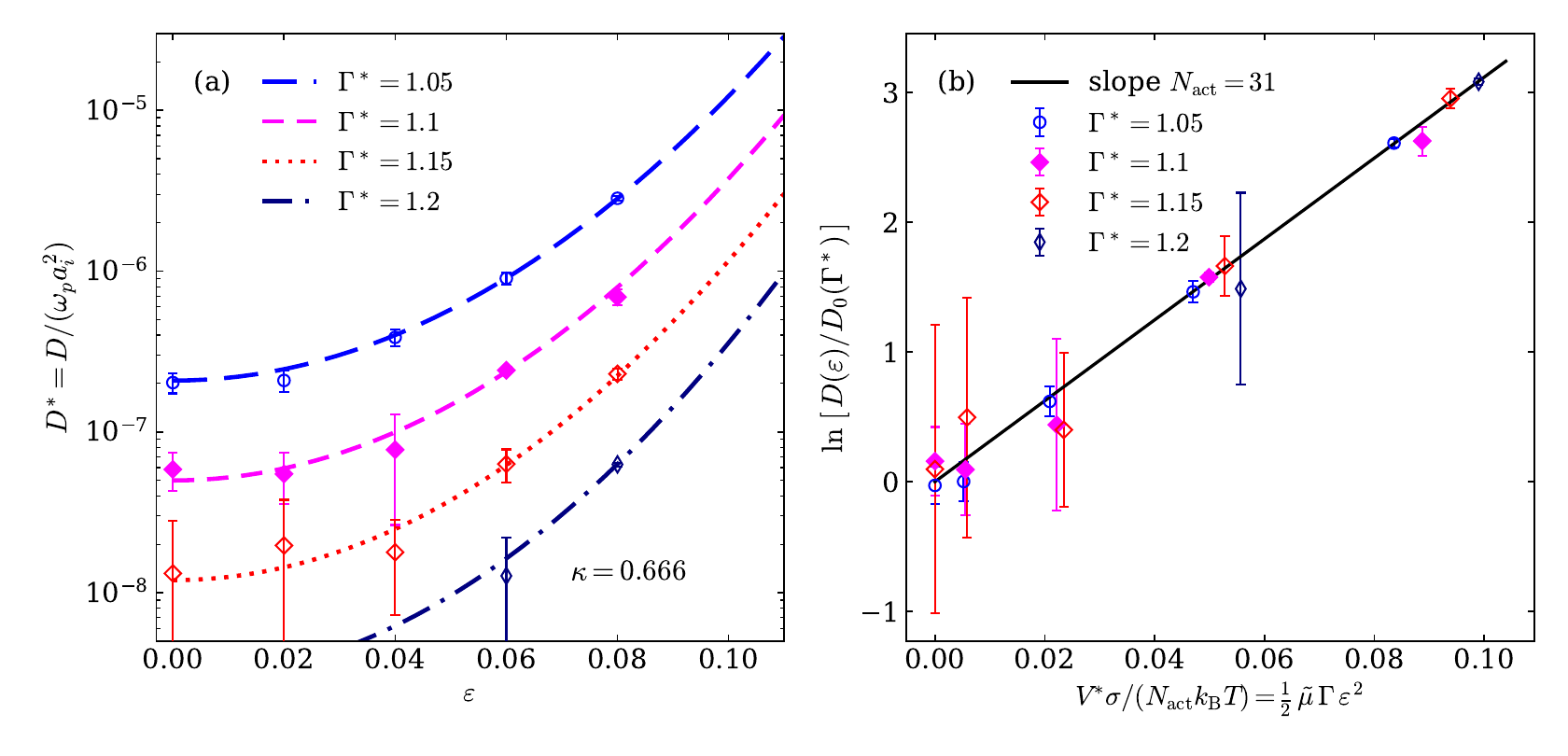}
  \caption{(a) Diffusion coefficients of the sheared bcc Yukawa crystal ($\kappa=0.666$, $16^3$ cells) versus shear strain $\varepsilon$, for $\Gamma^{*}=1.05$--$1.20$. Fit curves use  Eq.~(\ref{eq:master}) with the Caplan--Yaacoub form for $D_0(\Gamma^{*})$, refit to these runs,  multiplied by the stress factor $\exp(V^{*}\sigma/k_{\rm B}T)$ with a common activation volume $V^{*}$. (b) The same fit re-expressed using $\ln[D(\varepsilon)/D_0(\Gamma^{*})]$ versus the elastic energy stored per ion in units of $k_{\rm B}T$, Eq.~(\ref{eq:exponent}). All four couplings fall on one line through the origin whose slope is the number of ions in the activation volume, $N_{\rm act}=n_i V^{*}\simeq31$.}
  \label{fig:fit}
\end{figure*}

We now examine the lifetime of the cascades, however, we caution not to over-interpret measurements of the lifetime because they are dominated by the finite size of the simulation box rather than by any property of the crust. 
Once a Frenkel pair separates in the simulation the vacancy and interstitial diffuse independently and the number of hops in the intervening random walk is a three-dimensional quantity partially set by our box size. In a macroscopic volume, there would be some equilibrium density of defects constantly forming and annihilating, not necessarily with the partner they nucleated with. Our duty cycle here, generally being less than 1, therefore biases this quantity slightly relative to the thermodynamic limit. We present an analysis of cascade lengths in Fig.~\ref{fig:size}.
As for Fig.~\ref{fig:rate}, this analysis includes all cascades across all simulations where the duty cycle is less than 0.5 to avoid a counting bias in simulations where it becomes common that multiple Frenkel pairs are active at any given time, 
omitting thus only the simulations at $(\Gamma^*,\varepsilon)=(1.05, \,0.06)$ and $(1.05, \, 0.08)$.
We find a total of 1760 cascades in the remaining set, but restrict the analysis to the 1478 cascades with sizes of at least 10 nearest neighbor hops.

The simplest estimate is that a defect must diffuse of order one box crossing to find its partner across the periodic cell, requiring $\sim M^{2}$ hops for a cell of $M$ bcc units per side; with $M=16$ this is $\sim256$, matching the observed mean of $286$ hops (median 190, max 2804). We normalize assuming hops are to nearest neighbor sites using {$d_{\rm nn} = (3\pi^2)^{1/6} a_i\approx 1.76 a_i$} along the bcc body diagonal. The distributions do not show any apparent dependence on $\Gamma$ or $\varepsilon$, but this is unsurprising as the cascade size should be dominated by properties of the simulation box and simple random walk statistics that are common to every run. They are therefore flat in strain and in temperature. This analysis also indicates that the finite-size bias in the diffusion coefficient, indicated in Ref.\ \cite{Hughto2011Diffusion}, is primarily associated with the lifetime of cascades, not with formation of Frenkel pairs.

The distribution obtained by combining the data across strains in Fig.~\ref{fig:size}(b) is correspondingly broad with a tail to very large cascade sizes. There is possibly an excess at small cascades of about 10-15 hops. 
It is plausible that it is not noise, but a real second peak  associated with small closed loops, considered in Ref. \cite{Caplan2024}. Indeed, it is easy for the vacancy and interstitial to find each other again shortly after formation, before they become well separated.

In Fig.~\ref{fig:fit}(a) we show the measured $D^{*}(\Gamma^{*},\varepsilon)$ with fits. This is obtained from two-point fits ${D^{*} = (1/6)[{\rm MSD}(\tau) - {\rm MSD}(\epsilon)] / (\tau - \epsilon)}$ where $\tau$ is the total simulation time and $\epsilon$ is a small time after the start of the simulation so that $D^{*}$ is not biased upward by the thermal noise of particles on their lattice sites. 
We find a single activation volume
\begin{equation}
  N_{\rm act}\equiv n_i V^{*}= 31
  \qquad(\text{radius}\approx3\,a_i),
  \label{eq:fitvals}
\end{equation}
\noindent The single activation volume describes all
four to within the scatter. Fitting independently at each coupling gives 
$N_{\rm act}=32, \,28, \,32, \,37$ for $\Gamma^{*}=1.05,\,1.10,\,1.15,\,1.20$ respectively; stochasticity in the runs at lower temperatures and lower strains are largely responsible for the differences. The differences in best fit $N_{\rm act}$ are also partially affected by the fact the fit is with respect to $\varepsilon^2$ rather than $\varepsilon$ alone, causing the three lowest strains to cluster near the origin. This is shown in Fig.~\ref{fig:fit}(b), where plotting $\ln[D(\varepsilon)/D_0(\Gamma^{*})]$ against the elastic energy stored per ion in units of $k_{\rm B}T$, Eq.~(\ref{eq:exponent}), collapses every point onto one line through the origin whose slope is $N_{\rm act}$ as in Fig.~\ref{fig:rate}(a). With four temperatures and many seeds the data make the quadratic dependence obvious.

\begin{figure}[t]
  \includegraphics[width=\columnwidth]{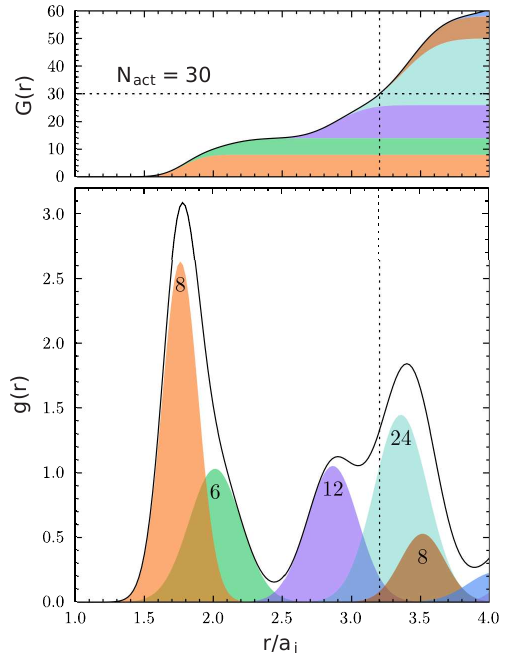}
  \caption{
  Schematic illustration of the radial distribution function $g(r)$ for the first few nearest neighbors in a bcc crystal.
  The colored peaks indicate the contribution of each neighbor-type, the number of neighbors of this type is also indicated.
  The lower panel is for $g(r)$, while the upper panel indicates integrated distribution function $G(r)$, i.e.\ average number of ions, located at distance less than $r$.
  The dotted lines indicate $G(r)=N_\mathrm{act}$ and respective $r$.
  For illustrative purposes the $g(r)$ was plotted using the harmonic lattice approximation for a Coulomb crystal at $\Gamma=180$ \cite{Baiko_ea00_HL,Chugunov_ea03_GR_HL}.}
  \label{fig:gr}
\end{figure}

This result has a simple qualitative interpretation. When $N_{\rm act}\simeq 31$ the activation volume $V^{*}=N_{\rm act}(4\pi a_i^{3}/3)$  has a radius of approximately $3a_i$. Comparing with radial distribution function $g(r)$ (see Fig.~\ref{fig:gr}), we conclude that $N_{\rm act}\simeq 31$ corresponds to 14 ions from the first peak (8 nearest neighbors at the corners of the unit and the 6 second nearest neighbors at the center of the adjacent unit cells), the 12 third nearest neighbors (located at the centers of the cells, bordered with the central cell by the edge; they contribute to the left shoulder of the second $g(r)$ peak) and the remaining $\sim 5$ ions can be attributed to the other ions of the second $g(r)$ peak, accounted with small  weight. This seems a reasonable estimate: a particle at the center of the activation volume must make a sufficiently large oscillation from its lattice site such that a nearest neighbor can move in to occupy its site, such that the original central particle becomes the interstitial with a vacancy left where the neighbor was. This likely involves several other nearest neighbors being at thermal maxima in their oscillations to lower the activation barrier and open up the physical volume for the incipient interstitial. 

\section{Discussion}

A static strain of a Yukawa crystal loads the lattice with elastic energy density $\tfrac12\mu\varepsilon^{2}$,
and the portion stored within the ${\sim}30$-ion nucleation region enhances the formation rate of vacancy--interstitial pairs. While this enhances the Arrhenius rate for defect formation it does not appreciably enhance the jump frequency of the vacancy or interstitial. At $\varepsilon=0.08$, roughly $80\%$ of the cold breaking strain, diffusion near the melting temperature is enhanced by an order of magnitude at astrophysically relevant screening lengths.

The model makes a prediction for other deformations that could also be checked with molecular dynamics in future work. Under a volume-conserving tensile stretch the relevant modulus is $(c_{11}-c_{12})/2$,
which for the bcc Yukawa crystal at $\kappa=0.666$ is $0.021\,n_i Z^{2}e^{2}/a_i$
\cite{Kozhberov2022}, a factor of ${\sim}8$ smaller than $C_{44}$. If $V^{*}$ is geometry independent, Eq.~(\ref{eq:master}) predicts the enhancement exponent to shrink by the same factor, rendering the effect nearly negligible at reachable strains. Simulations of stretched crystals can test this directly, but will require a much larger investment of computing time than was used in this work.

Because Eqs.~(\ref{eq:master})--(\ref{eq:kT}) are built from $\Gamma$ and
first-principles $\mu(\kappa,\Gamma)$, and because $\tilde\mu\Gamma_{\rm M}$ is nearly independent of $\kappa$ (Sec.~\ref{sec:theory}), the universal scaling of
Ref.~\cite{Caplan2024} carries over to the sheared crystal with nearly the same activation volume $V^{*}$ at every $\kappa$.
%AC: , as we have verified directly in
%Sec.~\ref{sec:kappa}.
%\ac{Sec.~\ref{sec:results}. 
The model can thus be applied across the crust and white dwarf core with a single parameter $\Gamma^*$. 
Thanks to the fact that $\tilde\mu\Gamma_{\rm M}$ is almost independent of the screening, the stress-induced enhancement of the diffusion can be described by a factor $\sim\exp[5(\varepsilon/0.1)^2\,\Gamma^\ast]$ when neglecting softening, leading to diffusion coefficient in strained Yukawa crystal:
\begin{equation}
    D(\Gamma^{*},\varepsilon) = 
     \frac{787}{\Gamma^{*}}
    \,\exp{ \left[ - \left( 20-5 \left(\frac{\varepsilon}{0.1} \right)^2 \right) \Gamma^{*} \right] },
\end{equation}
when combined with the universal fit \cite{Caplan2024}. 
Note, at high $\Gamma^{*}\gtrsim 1.5$ according to this estimate the diffusion stays very slow $D(\Gamma^{*})\lesssim 10^{-7}$ which is marginally detectable by MD simulations (e.g., \cite{Caplan2024}) even at critical deformation $\varepsilon\sim 0.1$ \cite{ChugunovHorowitz2010,Baiko2018}. Nevertheless, the stress-enhancement factor contributes strongly at higher $\Gamma^*$, so while it only provides an order of magnitude enhancement to diffusion near melting it can be many orders of magnitude at lower temperatures.

The astrophysical consequences are immediate. In magnetars, magnetic stresses drive the crust toward failure; stress-enhanced diffusion in the most strained regions opens a channel for thermally activated creep and gradual stress relaxation, shaping the approach to outbursts and low-amplitude oscillations \cite{Thompson2017,ChugunovHorowitz2010}. In accreting neutron stars, strains near the ocean--crust boundary would locally accelerate chemical separation \cite{Mckinven2016}, and in rotating neutron stars, enhanced diffusion in strained regions limits the longevity of elastically supported mountains relevant to continuous gravitational wave searches \cite{Ushomirsky2000}.

\begin{acknowledgments}

We thank Ashley Bransgrove, Nevin Smith, and Dany Yaacoub for discussion.  This work was supported by a grant from the Simons Foundation (MP-SCMPS-00001470) to MC. This research was supported in part by grant NSF PHY-2309135 to the Kavli Institute for Theoretical Physics (KITP). Financial support for this publication comes from Cottrell Scholar Award \#CS-CSA-2023-139 sponsored by Research Corporation for Science Advancement.
MC thanks the KITP for hospitality and MC acknowledges support as a KITP Scholar. 
The work of A.I.C. was funded by the baseline project FFUG-2027-0010 of the Ioffe Institute.
\end{acknowledgments}

%\bibliographystyle{apsrev4-2}
%\bibliography{references}

%apsrev4-2.bst 2019-01-14 (MD) hand-edited version of apsrev4-1.bst
%Control: key (0)
%Control: author (72) initials jnrlst
%Control: editor formatted (1) identically to author
%Control: production of article title (-1) disabled
%Control: page (0) single
%Control: year (1) truncated
%Control: production of eprint (0) enabled
%

\end{document}